\documentclass[review,12pt]{elsarticle}

\usepackage{amssymb}
\usepackage{amsthm}
\usepackage{framed} 
\usepackage{amsmath,color}
\usepackage{mathrsfs}
\usepackage{graphicx}
\usepackage{epstopdf}
\usepackage{float}
\usepackage{caption}
\usepackage{subcaption}
\usepackage{bm}
\usepackage{bbm}
\usepackage{mathrsfs}
\usepackage{cleveref}
\usepackage{soul}
\usepackage{accents}
\usepackage{color,soul} 
\usepackage{color} 
\usepackage{bm}
\usepackage{bbm}
\usepackage{upgreek} 
\usepackage{cprotect} 
\biboptions{sort&compress}
\soulregister\citep7 
\soulregister\citet7 
\soulregister\citealp7 
\newsavebox{\measurebox} 
\usepackage{titlesec} 
\usepackage[T1]{fontenc} 
\usepackage{lmodern} 
\journal{Philosophical Transactions of the Royal Society A}

\makeatletter
\def\@author#1{\g@addto@macro\elsauthors{\normalsize%
    \def\baselinestretch{1}%
    \upshape\authorsep#1\unskip\textsuperscript{%
      \ifx\@fnmark\@empty\else\unskip\sep\@fnmark\let\sep=,\fi
      \ifx\@corref\@empty\else\unskip\sep\@corref\let\sep=,\fi
      }%
    \def\authorsep{\unskip,\space}%
    \global\let\@fnmark\@empty
    \global\let\@corref\@empty  
    \global\let\sep\@empty}%
    \@eadauthor={#1}
}
\makeatother

\titleformat{\paragraph}
{\normalfont\normalsize\itshape}{\theparagraph}{1em}{}
\titlespacing*{\paragraph}
{0pt}{3.25ex plus 1ex minus .2ex}{1.5ex plus .2ex}

\begin{document}

\begin{frontmatter}



\title{An assessment of phase field fracture: crack initiation and growth}


\author{Philip K. Kristensen \fnref{DTU}}

\author{Christian F. Niordson\fnref{DTU}}

\author{Emilio Mart\'{\i}nez-Pa\~neda\corref{cor1}\fnref{ICL}}
\ead{e.martinez-paneda@imperial.ac.uk}

\address[DTU]{Department of Mechanical Engineering, Technical University of Denmark, DK-2800 Kgs. Lyngby, Denmark}

\address[ICL]{Department of Civil and Environmental Engineering, Imperial College London, London SW7 2AZ, UK}

\cortext[cor1]{Corresponding author.}

\begin{abstract}
The phase field paradigm, in combination with a suitable variational structure, has opened a path for using Griffith's energy balance to predict the fracture of solids. These so-called phase field fracture methods have gained significant popularity over the past decade, and are now part of commercial finite element packages and engineering fitness-for-service assessments. Crack paths can be predicted, in arbitrary geometries and dimensions, based on a global energy minimisation - without the need for \textit{ad hoc} criteria. In this work, we review the fundamentals of phase field fracture methods and examine their capabilities in delivering predictions in agreement with the classical fracture mechanics theory pioneered by Griffith. The two most widely used phase field fracture models are implemented in the context of the finite element method, and several paradigmatic boundary value problems are addressed to gain insight into their predictive abilities across all cracking stages; both the initiation of growth and stable crack propagation are investigated. In addition, we examine the effectiveness of phase field models with an internal material length scale in capturing size effects and the transition flaw size concept. Our results show that phase field fracture methods satisfactorily approximate classical fracture mechanics predictions and can also reconcile stress and toughness criteria for fracture. The accuracy of the approximation is however dependent on modelling and constitutive choices; we provide a rationale for these differences and identify suitable approaches for delivering phase field fracture predictions that are in good agreement with well-established fracture mechanics paradigms.
\end{abstract}


\end{frontmatter}


\section{Introduction}

It has been one hundred years since Alan Arnold Griffith \cite{Griffith1920} presented the energy balance that 
gave birth to the discipline of fracture mechanics. Cracks were postulated to propagate when the energy released due to crack growth is greater than or equal to the work required to create new free surfaces. Although this criterion for fracture is attractive, as it is based on simple thermodynamic principles, the fracture mechanics community soon moved in other directions to embrace local stress concepts such as stress intensity factors - a path opened by the work of Irwin \cite{Irwin1956}. More amenable to analytical and numerical solutions, these stress-intensity approaches came at the cost of imposing arbitrary criteria for determining the direction and extension of crack growth \cite{Anderson2005,Kendall2021}; as discussed below, the spatial and temporal evolution of crack paths are a natural byproduct of Griffith's energy balance. However, on the centenary of Griffith's seminal contribution, one can argue that the tables have been turned. The development of a variational stance for Griffith's theory and the subsequent pioneering use of the phase field paradigm to computationally track evolving cracks have again brought the view of fracture mechanics as an energetic problem in focus \citep{Bourdin2008}. Originating in the early 2000s but mainly developed over the past decade \citep{Bourdin2000,Miehe2010,Pons2010,Borden2012,Tanne2018}, the field of phase field fracture mechanics has enjoyed ever-increasing popularity up to its current ``quasi-hegemonic status'' \cite{Bourdin2019}.\\

The phase field fracture method has provided a suitable mathematical and computational framework for Griffith's energy balance. Phase field fracture analyses have proven capable of predicting - without \textit{ad-hoc} criteria - the nucleation, growth, merging, branching and arrest of cracks, in arbitrary dimensions and geometries (see, e.g., \cite{Borden2016,Miehe2016e,TAFM2021,Wu2021} and references therein). These capabilities are of increasing importance in advanced structural integrity assessment and the applications of phase field fracture have soared; examples include composite materials \cite{Quintanas-Corominas2019,CST2021}, shape memory alloys \cite{CMAME2021}, rock-like materials \cite{Zhou2018}, hydrogen embrittlement \cite{CMAME2018,CS2020}, functionally graded materials \cite{CPB2019,Kumar2021}, dynamic fracture \cite{Borden2012,McAuliffe2016}, fatigue damage \cite{Lo2019,Carrara2020}, ductile damage \citep{Miehe2016b,Alessi2018}, and Li-Ion batteries \cite{Miehe2015,Klinsmann2016c}. On the occasion of the fracture mechanics meeting organised at the Royal Society, and the associated Special Issue, we review the fundamentals of phase field fracture and gain new insight into its ability to deliver predictions in agreement with the classical fracture mechanics theory laid out by Griffith and his contemporaries.\\

The remainder of this paper is organised as follows. In Section \ref{sec:Theory} we introduce the phase field fracture theory, starting from Griffith's energy balance. The formulation is presented in a generalised fashion, accommodating any constitutive choice for the crack density function. The details of the numerical implementation are given in Section \ref{Sec:NumImplementation}, in the context of the finite element method. The main results and findings are presented in Section \ref{sec:Results}. First, we prescribe a remote $K$-field using a boundary layer model to quantify the energy released during crack initiation. Secondly, to investigate the capabilities of phase field fracture in accurately capturing stable crack growth, we use a double-cantilever beam with a known analytical solution for the crack extension as a function of the critical energy release rate and the applied load. Thirdly, using a plate of finite size with an edge crack, we investigate how phase field fracture models can capture size effects associated with the crack length. The present findings are discussed in the context of the literature in Section \ref{Sec:Discussion}. Finally, concluding remarks end the manuscript in Section \ref{Sec:Conclusions}.

\section{A variational framework for Griffith's energy balance}
\label{sec:Theory}

We shall describe the underlying mathematical formulation of phase field fracture models, focusing first on their construction as an approximation of Griffith's energy balance, and then present a generalised virtual work formulation in which the phase field is introduced as an additional independent kinematical descriptor. The theory refers to the response of a solid with volume $V$ occupying an arbitrary domain $\Omega \subset {\rm I\!R}^n$ $(n \in[1,2,3])$, with external boundary $\partial \Omega\subset {\rm I\!R}^{n-1}$, on which the outwards unit normal is denoted as $\mathbf{n}$.

\subsection{The phase field regularisation}

\begin{figure}[H]
  \makebox[\textwidth][c]{\includegraphics[width=1.2\textwidth]{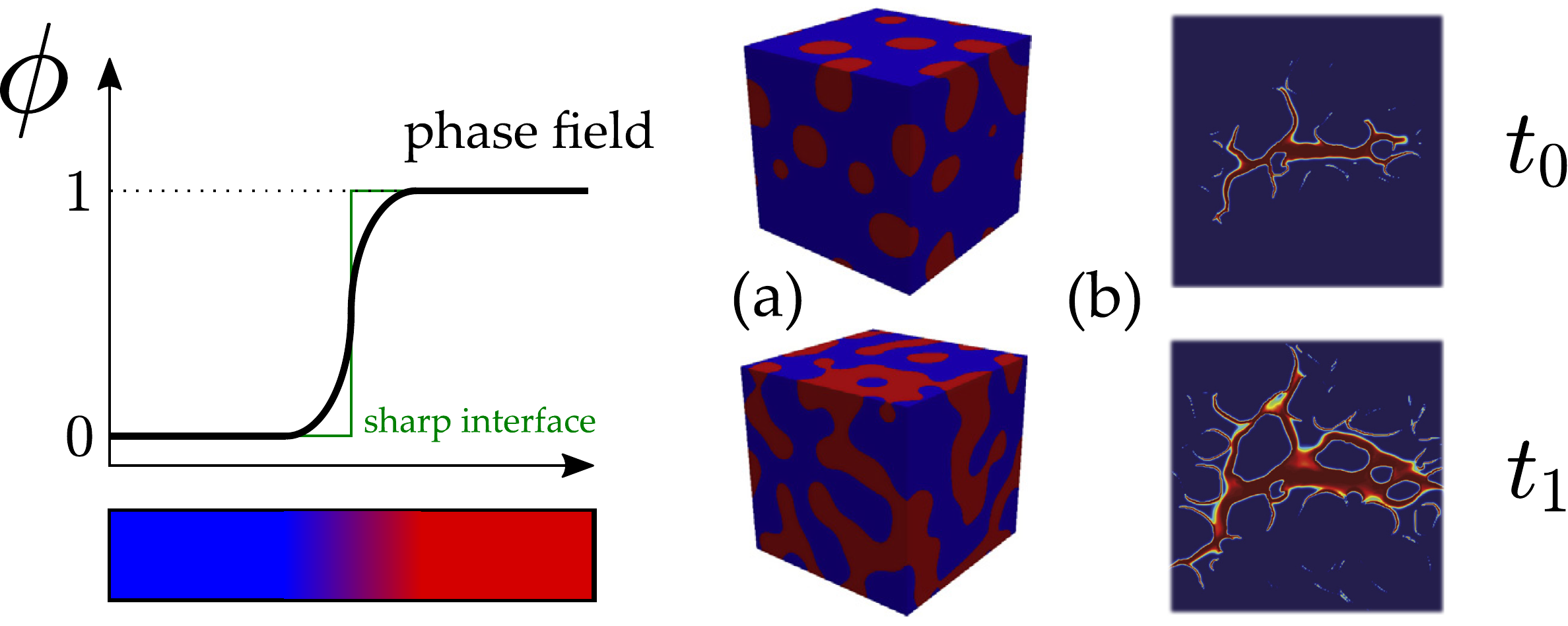}}%
  \caption{Tracking interfaces \emph{implicitly} using an auxiliary phase field $\phi$. Examples capturing (a) microstructural evolution \cite{Carolan2015}, and (b) the propagation of cracks \cite{Xia2017}, for two instants in time ($t_0$, $t_1$).}
  \label{fig:PhaseFieldIntro}
\end{figure}

From a continuum viewpoint, the first law of thermodynamics provides a detailed balance describing the interplay between the work done on the system, the internal energy, the kinetic energy, and the thermal power. Thus, thermodynamic equilibrium requires the total potential energy supplied by the internal strain energy density and external forces, $\Pi$, to remain constant. As noted by Griffith \cite{Griffith1920}, in the context of a fracture process under quasi-static and isothermal conditions, this entails balancing the reduction of potential energy that occurs during crack growth with the increase in surface energy resulting from the creation of new free surfaces. Mathematically, this can be formulated as follows. Consider a cracked solid with elastic strain energy $\Psi^e (\bm{\varepsilon})$ and elastic strain energy density $\psi^e (\bm{\varepsilon})$, which are a function of the strain tensor $\bm{\varepsilon}$. Under prescribed displacements, the variation of the total energy $\Pi$ due to an incremental increase in the crack area d$A$ is given by
\begin{equation}
\frac{\text{d} \Pi}{\text{d} A} = \frac{\text{d} \Psi^e (\bm{\varepsilon})}{\text{d} A} + \frac{\text{d} W_c}{\text{d} A}  = 0,
\end{equation}

\noindent where $W_c$ is the work required to create new surfaces. The last term is the so-called critical energy release rate $G_c=\text{d} W_c / \text{d} A$, a material property that characterises the fracture resistance of the solid. Therefore, a pre-existing crack will grow as soon as the elastic energy stored in the material $\psi^e$ is sufficiently large to overcome the material toughness $G_c$. Griffith's energy balance can be formulated in a variational form as \cite{Francfort1998}:
\begin{equation}\label{Eq:Pi}
\Pi = \int_\Omega \psi^e \left( \bm{\varepsilon} \right) \text{d} V + \int_\Gamma   G_c \, \text{d} \Gamma,
\end{equation}

\noindent where $\Gamma$ is the crack surface. Griffith's minimality principle is now global and cracking phenomena can be captured by minimising (\ref{Eq:Pi}), with crack behaviour (nucleation, trajectory, etc.) being dictated \emph{only} by the exchange between elastic and fracture energies. However, minimisation of (\ref{Eq:Pi}) is hindered by the unknown nature of $\Gamma$, making the problem computationally intractable. This obstacle can be addressed by exploiting the phase field paradigm - an auxiliary (phase) field variable $\phi$ can be defined to describe discrete discontinuous phenomena, such as cracks, with a smooth function. As illustrated in Fig. \ref{fig:PhaseFieldIntro}, the key idea is to smear a sharp interface into a \emph{diffuse} region using this phase field order parameter $\phi$, which takes a distinct value for each of the two phases (e.g., 0 and 1) and exhibits a smooth change between these values near the interface. The use of phase field variables to \emph{implicitly} track interfaces has gained significant traction in the condensed matter and materials science communities, becoming the most widely used technique for modelling microstructural evolution \cite{Provatas2011}. Also, the success has been recently extended to the phenomenon of corrosion, where the phase field is used to describe the solid metal - aqueous electrolyte interface \cite{JMPS2021}.

In the context of fracture mechanics, the phase field can be used to track the solid-crack interface, enabling the handling of cracks with arbitrary topological complexity, as well as their potential interactions. Thus, the phase field resembles a damage variable, taking (e.g.) the value of $\phi=0$ in intact regions and of $\phi=1$ in fully cracked material points. Equally important, the evolution law for the phase field variable is grounded on Griffith's energy balance. Accordingly, the Griffith functional (\ref{Eq:Pi}) can be approximated using the following phase field-regularised functional: 
\begin{equation}\label{Eq:Piphi}
\Pi_\ell = \int_\Omega \big[ g \left( \phi \right) \psi_0^e \left( \bm{\varepsilon} \right) + G_c \gamma \left( \phi, \ell \right) \big] \,  \text{d} V
\end{equation}

\noindent where $\ell$ is a length scale parameter that governs the size of the fracture process zone, $\psi_0^e$ denotes the elastic strain energy density of the undamaged solid, and $\gamma$ is the so-called crack surface density function \cite{Miehe2010}. The work required to create a cracked surface is now expressed as a volume integral, making the problem computationally tractable. Also, a degradation function $g \left( \phi \right)$ is defined following continuum damage mechanics arguments, such that the stiffness of the solid is degraded as the phase field approaches the value corresponding to the crack phase (e.g., $\psi^e =0$ for $\phi=1$). Choices for crack surface density function $\gamma$ have been mostly inspired in the Ambrosio and Tortorelli \cite{Ambrosio1990} approximation of the Mumford-Shah potential \cite{Mumford1989} - a well-known functional in image segmentation that closely resembles the variational fracture formulation described here. Upon these constitutive choices for $\gamma$, it can be proven using $\Gamma$-convergence that the regularised functional $\Pi_\ell$ (\ref{Eq:Piphi}) converges to the Griffith functional $\Pi$ (\ref{Eq:Pi}) when $\ell \to 0^+$ \cite{Bellettini1994,Chambolle2004}. Thus, $\ell$ can be interpreted as a regularising parameter in its vanishing limit. However, for $\ell>0^+$ a finite material strength is introduced and thus $\ell$ becomes a material property governing the strength \citep{Tanne2018}; e.g., for plane stress:
\begin{equation}\label{eq:CriticalStress}
  \sigma_c  \propto \sqrt{\frac{G_c E}{\ell}} = \frac{K_{Ic}}{\sqrt{\ell}}
\end{equation}
\noindent where $K_{Ic}$ is the material fracture toughness and $E$ denotes Young's modulus. From a numerical perspective, the presence of a length scale $\ell$ regularises the problem, ensuring mesh-objectivity as the model is non-local. We conclude this part by emphasising that Eq. (\ref{Eq:Piphi}) provides a rigorous approximation to Griffith's energy balance that is amenable to numerical computations. Fracture can be predicted with no other consideration than the minimisation of a free energy functional composed of the stored elastic bulk energy plus the fracture energy.

\subsection{Principle of virtual work. Balance of forces}

The primal kinematic variables of the model are the displacement field $\mathbf{u}$ and the damage phase field $\phi$. We restrict our attention to small strains and isothermal conditions. Accordingly, the strain tensor $\bm{\varepsilon}$ is given by
\begin{equation}
    \bm{\varepsilon} = \frac{1}{2}\left(\nabla\mathbf{u}^T+\nabla\mathbf{u}\right) \, .
\end{equation}

The balance equations for the coupled deformation-fracture system are now derived using the principle of virtual work. We use $\delta$ to denote virtual quantities and introduce the Cauchy stress $\bm{\sigma}$, which is work conjugate to the strains $\bm{\varepsilon}$. Accordingly, a traction $\mathbf{T}$ is defined, which is work conjugate to the displacements $\mathbf{u}$. Regarding damage, we introduce a scalar stress-like quantity $\omega$, which is work conjugate to the phase field $\phi$, and a phase field micro-stress vector $\bm{\upxi}$ that is work conjugate to the gradient of the phase field $\nabla\phi$. The phase field is assumed to be driven by the displacement problem alone. As a result, no external traction is associated with $\phi$. Accordingly, in the absence of body forces, the principle of virtual work reads: 
\begin{equation}
 \int_\Omega \big\{ \bm{\sigma}:\delta\bm{\varepsilon}  + \omega\delta\phi+\bm{\upxi} \cdot \delta \nabla \phi
    \big\} \, \text{d}V =  \int_{\partial \Omega} \left( \mathbf{T} \cdot \delta \mathbf{u} \right) \, \text{d}S
\end{equation}
\noindent This equation must hold for an arbitrary domain $\Omega$ and for any kinematically admissible variations of the virtual quantities. Thus, by making use of the fundamental lemma of the calculus of variations, the local force balances are given by: 
\begin{equation}
    \begin{split}
        &\nabla\cdot\bm{\sigma}=0  \\
        &\nabla\cdot\bm{\upxi}-\omega =0
    \end{split}\hspace{2cm} \text{in } \Omega,\label{eq:balance}
\end{equation}

\noindent with natural boundary conditions: 
\begin{equation}
    \begin{split}
        \bm{\sigma}\cdot\mathbf{n}=\mathbf{T} \\
         \bm{\upxi} \cdot \mathbf{n}=0 
    \end{split} \hspace{2cm} \text{on } \partial\Omega.\label{eq:balance_BC}
\end{equation}

\subsection{Constitutive theory}

The potential energy density of the solid is defined as,
\begin{equation}
    \psi \left( \bm{\varepsilon}, \, \phi, \, \nabla \phi \right) = \psi^e + \psi^f = g \left( \phi \right) \frac{1}{2} \bm{\varepsilon} : \bm{C}_0 : \bm{\varepsilon} + G_c \gamma \left( \phi, \, \nabla \phi \right) \, .
\end{equation}

\noindent Here, $\psi^f$ is the fracture energy and $\bm{C}_0$ is the linear elastic stiffness matrix. Accordingly, the Cauchy stress tensor is derived as,
\begin{equation}
    \bm{\sigma} = \frac{\partial \psi}{\partial \bm{\varepsilon}} = g \left( \phi \right) \left( \bm{C}_0 : \bm{\varepsilon} \right) \, .
\end{equation}

We shall now proceed to make constitutive choices for the phase field fracture formulation. The two models that are arguably most widely used will be considered and the implications of these constitutive choices investigated. First, we note that the degradation function $g \left( \phi \right)$ should be continuous and monotonic, and take the values $g(0)=1$ and $g(1)=0$; the following quadratic form is adopted,
\begin{equation}
    g \left( \phi \right) = \left( 1 - \phi \right)^2 \, .
\end{equation}

Secondly, restricting our attention to phase field formulations derived from the family of Ambrosio-Tortorelli functionals, we proceed to define the crack surface density function $\gamma \left( \phi \right)$ and the crack surface $A$ as follows:
\begin{equation}\label{eq:McMeeking}
    A = \int_\Omega \gamma \left( \phi \right) \, \text{d} V =  \int_\Omega  \frac{1}{4c_w\ell}\left( w(\phi) + \ell^2 |\nabla\phi|^2\right) \, \text{d} V \, .
\end{equation}

\noindent Here, the function $w(\phi)$ must fulfill $w(0)=0$ and $w(1)=1$, and
\begin{equation}
    c_w = \int_0^1\sqrt{w(\varphi)} \, \text{d}\varphi \, .
\end{equation}

\begin{figure}[H]
    \centering
    \includegraphics[width=0.8\textwidth]{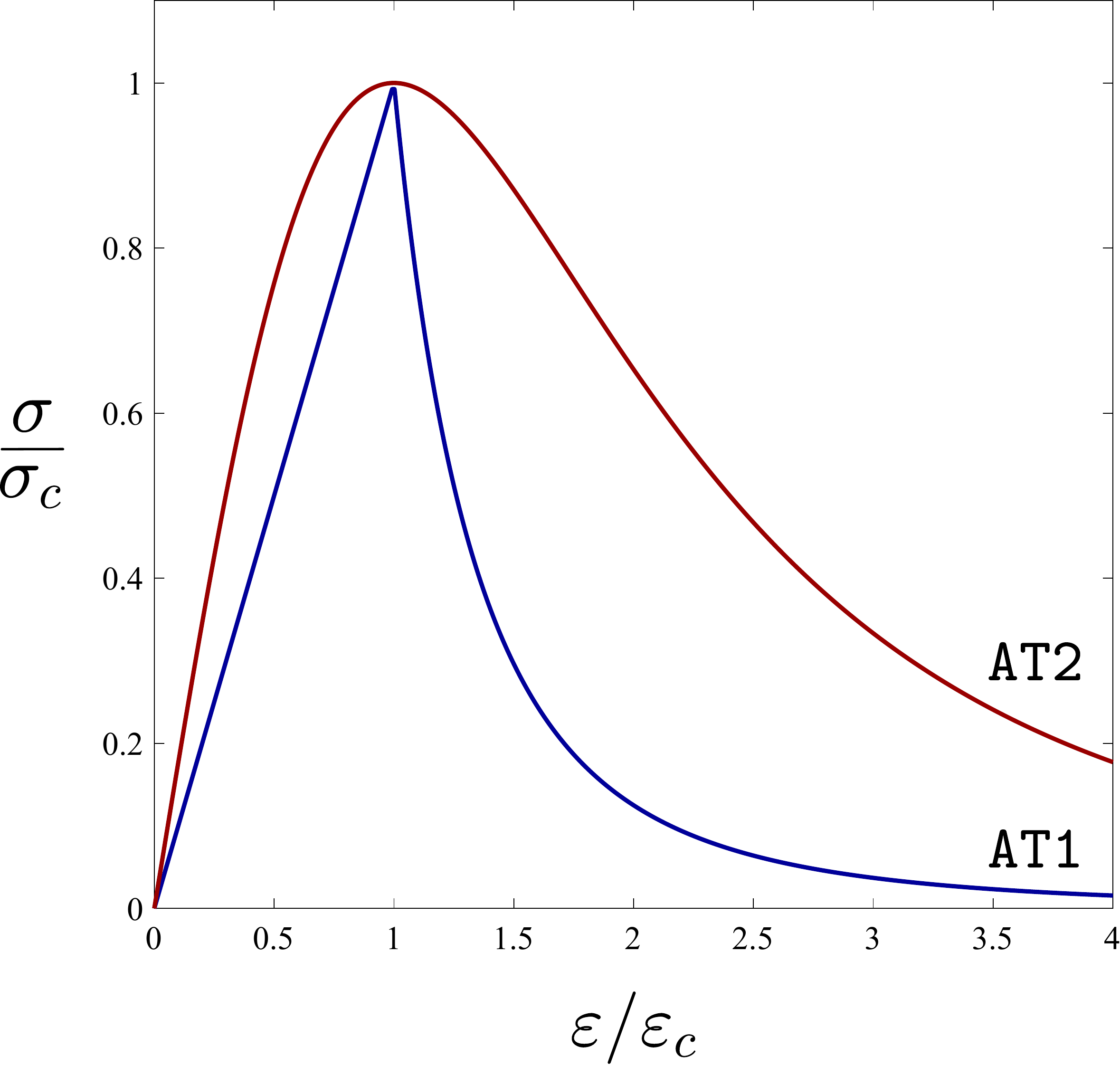}
    \cprotect\caption{Uniaxial stress-strain responses predicted by the \verb|AT1| and \verb|AT2| constitutive choices of the crack density function.}
    \label{fig:AT1AT2tractionseparation}
\end{figure}

The choice of $w (\phi )=\phi^2$ ($c_w=1/2$) renders the so-called \emph{standard} or \verb|AT2| phase field model \cite{Ambrosio1990}, while the choice $w (\phi )=\phi$ ($c_w=2/3$) introduces an elastic regime prior to the onset of damage, and is often referred to as the \verb|AT1| model \cite{Pham2011}. The stress-strain response resulting from the solution to the homogeneous 1D problem ($\nabla \phi=0$) is shown in Fig. \ref{fig:AT1AT2tractionseparation} for both models. It can be readily seen how the \verb|AT1| model exhibits a linear response until reaching the critical stress, while the \verb|AT2| results deviate earlier from the undamaged stress-strain response. Also, the \verb|AT1| model exhibits a sharper drop of the stress upon reaching the material strength. The critical failure stress attained for each model is given by \cite{CMAME2018}:
\begin{equation}\label{eq:sigmacAT1AT2}
   \sigma^{\verb|AT1|}_c=\sqrt{\dfrac{3EG_c}{8\ell}}, \hspace{2cm} \sigma^{\verb|AT2|}_c=\dfrac{3}{16}\sqrt{\dfrac{3EG_c}{\ell}}
\end{equation}

\noindent Thus, as $\ell \to 0$, the material strength goes to infinity; this is consistent with linear elastic fracture mechanics and $\Gamma$-convergence arguments. At this point, it should be noted that many other constitutive choices have been proposed in the literature. For example, some models are based on the Ginzburg-Landau formulations used in phase transition studies \cite{Karma2001}, while others, closer to the formulation presented here, aim at coupling phase field with cohesive zone concepts \cite{Freddi2017,Wu2017,Wu2018a}.

The fracture micro-stress variables $\omega$ and $\bm{\upxi}$, which can have energetic and dissipative parts, are defined as follows. Independently of the constitutive choices outlined above (\verb|AT1| vs \verb|AT2|), we derive the scalar micro-stress as,
\begin{equation}\label{eq:consOmega}
    \omega = \dfrac{\partial\psi}{\partial\phi} = {g^{\prime}(\phi)}\psi^e+\frac{1}{4c_w \ell} G_{\mathrm{c}} w^{\prime}(\phi) \, .
\end{equation}

\noindent Similarly, the phase field micro-stress vector $\bm{\upxi}$ reads:
\begin{equation}\label{eq:consXi}
    \bm{\upxi} = \dfrac{\partial\psi}{\partial\nabla\phi} = \frac{\ell}{2c_w} G_{\mathrm{c}} \nabla \phi \, .
\end{equation}

The phase field evolution law (\ref{eq:balance})b can then be reformulated accordingly,
\begin{equation}\label{eq:PhaseFieldStrongForm}
  \frac{G_c}{2c_w\ell}   \left( \frac{w^{\prime}(\phi)}{2} - \ell^2 \nabla^2 \phi \right) + {g^{\prime}(\phi)}\psi^e \left( \bm{\varepsilon} \right) = 0  
\end{equation}

\noindent showcasing the competition between the stored elastic energy and the fracture energy.

\section{Numerical implementation}
\label{Sec:NumImplementation}

We proceed to describe the numerical implementation, in the context of the finite element method, of the variational fracture framework described in Section \ref{sec:Theory}. First, we introduce a history field variable to ensure damage irreversibility. Secondly, we address the discretisation of the weak formulation, formulate the residuals and the stiffness matrices, and discuss solution schemes for the two-field problem. The implementation is conducted within an Abaqus user-element (UEL) subroutine, with the pre-processing of the input files carried out using Abaqus2Matlab \citep{AES2017}.

\subsection{Damage irreversibility}
\label{Sec:DamageIrreversibility}

Following Miehe \textit{et al.} \cite{Miehe2010}, a history variable field $\mathcal{H}$ is introduced to prevent crack healing, ensuring that the following condition is always met
\begin{equation} \label{growth}
    \phi_{t+\Delta t} \geq \phi_{t} \, ,
    \centering
\end{equation}

\noindent where $\phi_{t+\Delta t}$ is the phase field variable in the current time step while $\phi_{t}$ denotes the value of the phase field on the previous time step. For both loading and unloading scenarios, the history field must satisfy the Kuhn-Tucker conditions:
\begin{equation}
    \psi_{0}^e - \mathcal{H} \leq 0 \text{,} \hspace{7mm} \dot{\mathcal{H}} \geq 0 \text{,} \hspace{7mm} \dot{\mathcal{H}}(\psi_{0}^e-\mathcal{H})=0 \, .
    \centering
\end{equation}
\noindent Accordingly, the history field for a current time $t$ over a total time $\tau$ can be written as: 
\begin{equation}\label{eq:History}
     \mathcal{H} = \max_{\tau \in[0,t]}\psi_0^e( \tau). 
\end{equation}

\subsection{Finite element discretisation}
\label{subsec:DisFEM}

We shall now describe the finite element discretisation. Our numerical implementation uses as nodal unknowns the displacement vector $\hat{\mathbf{u}}$ and the phase field $\hat{\phi}$ fields. Considering the history field $\mathcal{H}$ described above and the constitutive choices outlined in Section \ref{sec:Theory}, one can formulate the weak form of the two-field problem as,
\begin{equation}\label{eq:weak}
    \int_\Omega \left\{ \big[ g \left( \phi \right) + \kappa \big] \bm{\sigma}_0 : \delta \bm{\varepsilon} - g^{\prime} \left( \phi \right) \mathcal{H} \delta \phi + \frac{G_c}{2c_w \ell}   \left( \frac{w^{\prime}(\phi)}{2} \delta \phi - \ell^2 \nabla \phi \nabla \delta \phi \right) \right\} \text{d} V = 0  \, .
\end{equation}

\noindent Here, $\bm{\sigma}_0$ is the undamaged Cauchy stress tensor and $\kappa$ is a small positive-valued constant that is introduced to prevent ill-conditioning when $\phi=1$; a value of $\kappa=1 \times 10^{-7}$ is here adopted.  

Making use of Voigt notation, the nodal quantities are interpolated as: 
\begin{equation}\label{eq:Ndiscret}
\mathbf{u} = \sum_{i=1}^m \bm{N}_i \hat{\mathbf{u}}_i, \hspace{2.5cm} \phi =  \sum_{i=1}^m N_i \hat{\phi}_i \, ,
\end{equation}
\noindent where $m$ is the total number of nodes per element, $N_i$ denotes the shape function associated with node $i$ and $\bm{N}_i$ is the shape function matrix, a diagonal matrix with $N_i$ in the diagonal terms. Similarly, the associated gradient quantities can be discretised using the corresponding $\bm{B}$-matrices, containing the derivative of the shape functions, such that:
\begin{equation}\label{eq:Bdiscret}
\bm{\varepsilon} = \sum\limits_{i=1}^m \bm{B}^{\bm{u}}_i \hat{\mathbf{u}}_i, \hspace{2.5cm}  \nabla\phi =  \sum\limits_{i=1}^m \mathbf{B}_i \hat{\phi}_i \, .
\end{equation}

Considering the weak form (\ref{eq:weak}) and the discretisation (\ref{eq:Ndiscret})-(\ref{eq:Bdiscret}), we derive the residuals for each primal kinematic variable as:
\begin{align}
    & \mathbf{R}_i^\mathbf{u} = \int_\Omega \left\{ \big[ g \left( \phi \right) + \kappa \big] \left(\bm{B}^\mathbf{u}_i\right)^T \bm{\sigma}_0 \right\} \, \text{d}V \, , \\
    & R_i^\phi = \int_\Omega \left\{ g^{\prime} \left( \phi \right) N_i \mathcal{H}+ \frac{G_c}{2c_w \ell}  \left[\frac{w^{\prime}(\phi)}{2} N_i + \ell^2 \,  \left( \mathbf{B}_i \right)^T \nabla\phi\right]\right\}dV \,.
\end{align}

\noindent And finally, the consistent tangent stiffness matrices $\bm{K}$ are then obtained as follows:
\begin{align}
    & \bm{K}_{ij}^{\mathbf{u}} = \frac{\partial \bm{R}_{i}^{\bm{u}} }{\partial \bm{u}_{j} } = 
        \int_{\Omega} \left\{ \big[ g \left( \phi \right) + \kappa \big] {(\bm{B}_{i}^{\bm{u}})}^{T} \bm{C}_0 \, \bm{B}_{j}^{\bm{u}} \right\} \, \text{d}V \, , \\
    & \bm{K}_{ij}^{\phi} = \frac{\partial R_{i}^{\phi} }{ \partial \phi_{j} } =  \int_{\Omega} \left\{ \left( {g^{\prime \prime}(\phi)} \mathcal{H} + \frac{G_{c}}{4 c_w \ell}  {w^{\prime \prime}(\phi)} \right) N_{i} N_{j} + \frac{G_{c} \ell}{2 c_w} \, \mathbf{B}_i^T\mathbf{B}_j \right\} \, \text{d}V \, .
\end{align}
Therefore, the global system of equations reads,
\begin{equation}\label{Eq:GlobalElementSystem}
    {\begin{Bmatrix}
        \textbf{u}\\[0.3em] \bm{\phi}
    \end{Bmatrix}}_{t+\Delta t} = 
    {\begin{Bmatrix}
        \textbf{u}\\[0.3em] \bm{\phi}
    \end{Bmatrix}}_{t} -
    {\begin{bmatrix}
        \mathbf{K}^{\mathbf{u}}& 0 \\[0.3em] 
        0 & \mathbf{K}^{\phi}
    \end{bmatrix}}_{t}^{-1}
    {\begin{Bmatrix}
        \textbf{R}^{\textbf{u}}\\[0.3em] \textbf{R}^{\phi}
    \end{Bmatrix}}_{t}
    \centering
\end{equation}

Several schemes have been proposed to obtain the solutions for which \(\textbf{R}^{\textbf{u}}=\bm{0}\) and \(\textbf{R}^{\phi}=\bm{0}\). In so-called \emph{monolithic} solution schemes, the displacement and phase field sub-systems are solved simultaneously; while \emph{staggered} (or alternate minimisation) approaches solve each sub-system sequentially. Monolithic solution strategies are unconditionally stable and, therefore, more efficient. However, the total potential energy functional (\ref{Eq:Piphi}) is non-convex with respect to $\mathbf{u}$ and $\phi$, hindering convergence. Contrarily, for a fixed $\mathbf{u}$, Eq. (\ref{Eq:Piphi}) is convex with respect to $\phi$ (and vice-versa) and the associated robustness has made staggered solution schemes more popular. Notwithstanding, it has been recently demonstrated that the use of quasi-Newton methods such as the Broyden-Fletcher-Goldfarb-Shanno (BFGS) algorithm enables the implementation of robust monolithic schemes that are very efficient and do not exhibit convergence issues 
\citep{Wu2020a,TAFM2020}. Accordingly, the BFGS algorithm is employed here, in conjunction with a monolithic approach. 

\section{Results}
\label{sec:Results}

We shall now model three paradigmatic boundary value problems to investigate the capabilities of phase field fracture models in predicting crack initiation and growth in agreement with the fracture energy balance. First, in Section \ref{Sec:BoundaryLayer}, the onset of crack growth is investigated by applying a remote energy release rate through a boundary layer model. Secondly, we model crack propagation in a double cantilever beam to compare phase field predictions with analytical results derived from beam theory from a known applied displacement and material toughness (Section \ref{Sec:DCB}). Finally, in Section \ref{Sec:TransitionFlawSize}, we show how size effects and the transition flaw size concept are a natural byproduct of phase field fracture models.

\subsection{Initiation of crack growth: prescribing a remote $G$}
\label{Sec:BoundaryLayer}

\begin{figure}[H]
    \centering
    \includegraphics[width=0.8\textwidth]{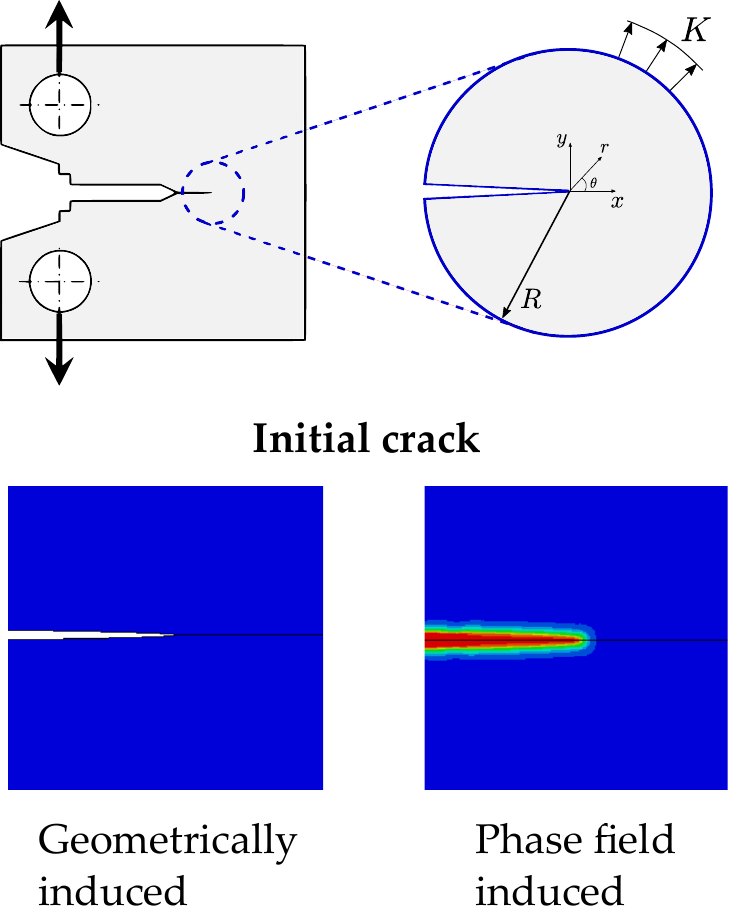}
    \caption{Boundary layer analysis: illustration of the boundary value problem and of the approaches adopted to introduce the initial crack in the solid.}
    \label{fig:MBLSketch}
\end{figure}

The initiation of crack growth is investigated under plane strain conditions using a so-called boundary layer model. As illustrated in Fig. \ref{fig:MBLSketch}, the crack tip fields can be characterised as a function of the remote elastic $K$-field. Thus, considering a polar coordinate system ($r, \theta$) and a Cartesian coordinate system ($x, y$) centred at the crack tip, with the crack plane along the negative $x$-axis, the displacement is given by the first term in Williams expansion \cite{Williams1957}:
\begin{equation}\label{eq:Williams}
    u_i = \frac{K}{E} r^{1/2} f_i \left( \theta, \nu \right) \, ,
\end{equation}

\noindent where the subscript $i$ denotes the Cartesian components, and the functions $f_i ( \theta, \nu )$ are given by,
\begin{equation}
    f_x = \frac{1+\nu}{\sqrt{2 \pi}} \left( 3 - 4 \nu - \cos \theta \right) \cos \left( \frac{\theta}{2} \right) \, ,
\end{equation}
\begin{equation}\label{eq:Williams1}
    f_y = \frac{1+\nu}{\sqrt{2 \pi}} \left( 3 - 4 \nu - \cos \theta \right) \sin \left( \frac{\theta}{2} \right) \, .
\end{equation}

\noindent Here, $\nu$ is Poisson's ratio. The relationship between the stress intensity factor $K$ and the energy release rate $G$ is given by Irwin's relation \cite{Irwin1956}:
\begin{equation}\label{eq:Irwin}
    G= \left( 1 - \nu^2 \right) \frac{K^2}{E} \, .
\end{equation}

Consequently, the crack tip mechanics for a given remote $G$ (or $K$) can be evaluated by prescribing the displacements of the nodes located in the outer boundary of the finite element model following (\ref{eq:Williams})-(\ref{eq:Irwin}). Only one half of the boundary layer geometry is modelled due to symmetry. The model is discretised using approximately 30,000 quadratic quadrilateral elements with reduced integration. The mesh is refined in the crack propagation region, where the element aspect ratio is kept equal to 1. Throughout this manuscript, Poisson's ratio is given by $\nu=0.3$.

The initiation of crack growth is investigated considering both the \verb|AT1| and \verb|AT2| constitutive choices for the crack density function. Also, we assess the influence of the two approaches that can potentially be used to define the initial crack: (1) \textit{via} the phase field, by defining $\phi=1$ as the initial condition (or enforcing $\mathcal{H} \to \infty$), and (2) geometrically, by duplicating the nodes along the crack faces; see Fig. \ref{fig:MBLSketch}. The aim is to assess whether phase field fracture models predict the initiation of cracking at $G=G_c$, as it would be expected based on the classical fracture mechanics theory. Fracture is unstable, exhibiting a flat crack growth resistance response and therefore the remote load at initiation (as characterised by $G$ or $K$) can be easily identified. By dimensional analysis, the length scales governing the problem are the phase field length scale $\ell$, the element size $h$ and a fracture or characteristic material length (see, e.g. \cite{Falk2001a}):
\begin{equation}
    L_f = \frac{G_c (1-\nu^2)}{E} \, ;
\end{equation}

\noindent Note that the initial crack length $a$ and the outer radius of the boundary layer $R$ (chosen such that $R,a>>\ell, L_f$) are not relevant in the present boundary value problem. Thus, we investigate the role that the two remaining non-dimensional groups ($\ell/h$, $L_f/\ell$) play on our fracture mechanics assessment.\\

\begin{figure}[H]
    \centering
    \includegraphics[width=\textwidth]{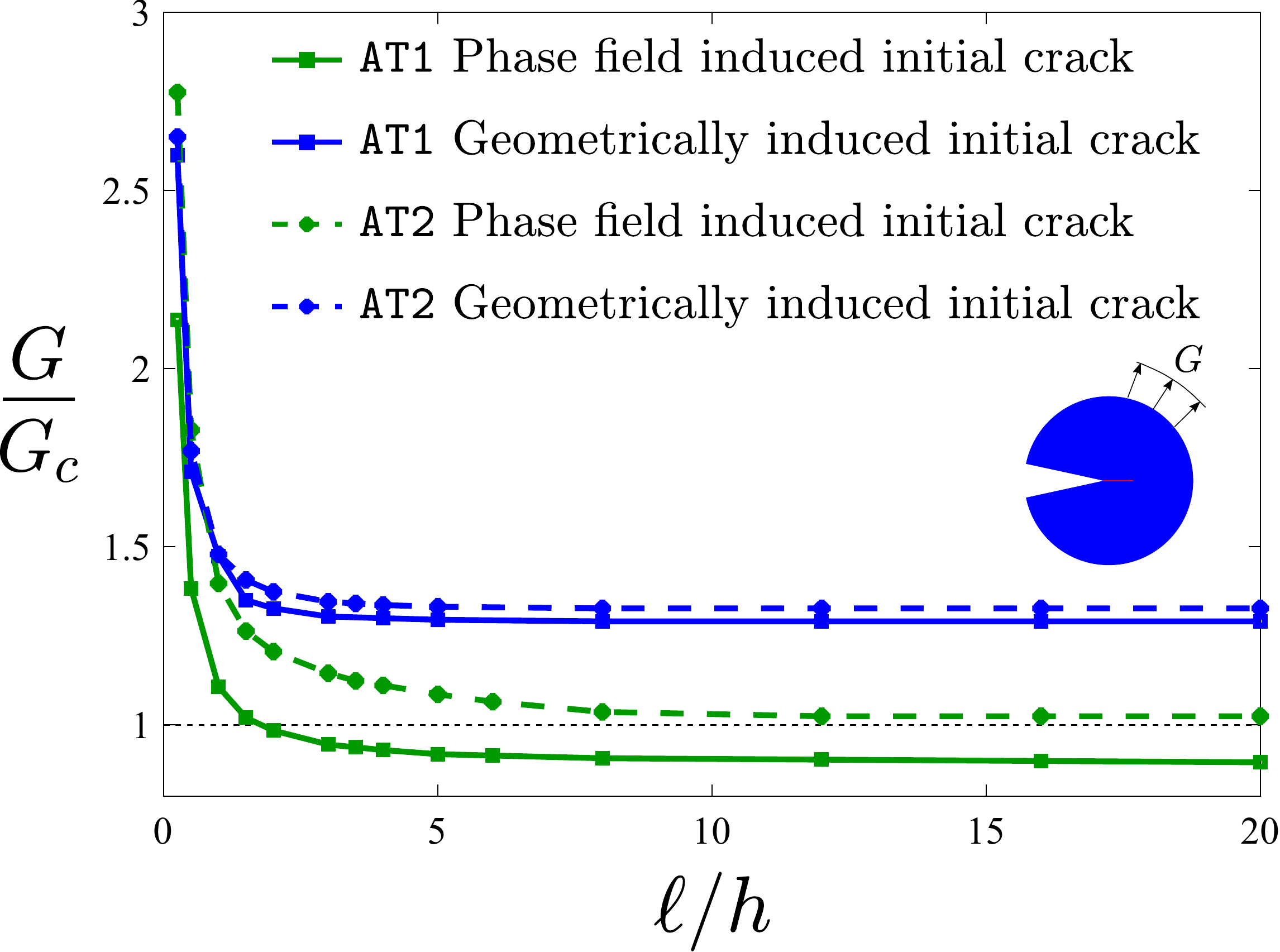}
    \caption{Boundary layer analysis: Mesh sensitivity of the energy release rate at initiation of crack growth for different constitutive choices and approaches for implementing the initial crack. The value of the characteristic fracture length scale to the phase field length scale is $L_f/\ell=10$.}
    \label{fig:LengthScales}
\end{figure}

The results obtained for the non-dimensional group $\ell/h$ are shown in Fig. \ref{fig:LengthScales}. Note first that, in all cases, the solution appears to converge when the mesh sufficiently resolves the phase field and fracture length scales. The results are essentially identical if eight elements or more are used to resolve $\ell$. This is not unexpected given the mesh objectivity of non-local models but element length-dependent corrections for the surface energy have been proposed \cite{Bourdin2008}\footnote{No mesh anisotropy effects are investigated in this work.}.

\begin{figure}[H]
  \makebox[\textwidth][c]{\includegraphics[width=1.25\textwidth]{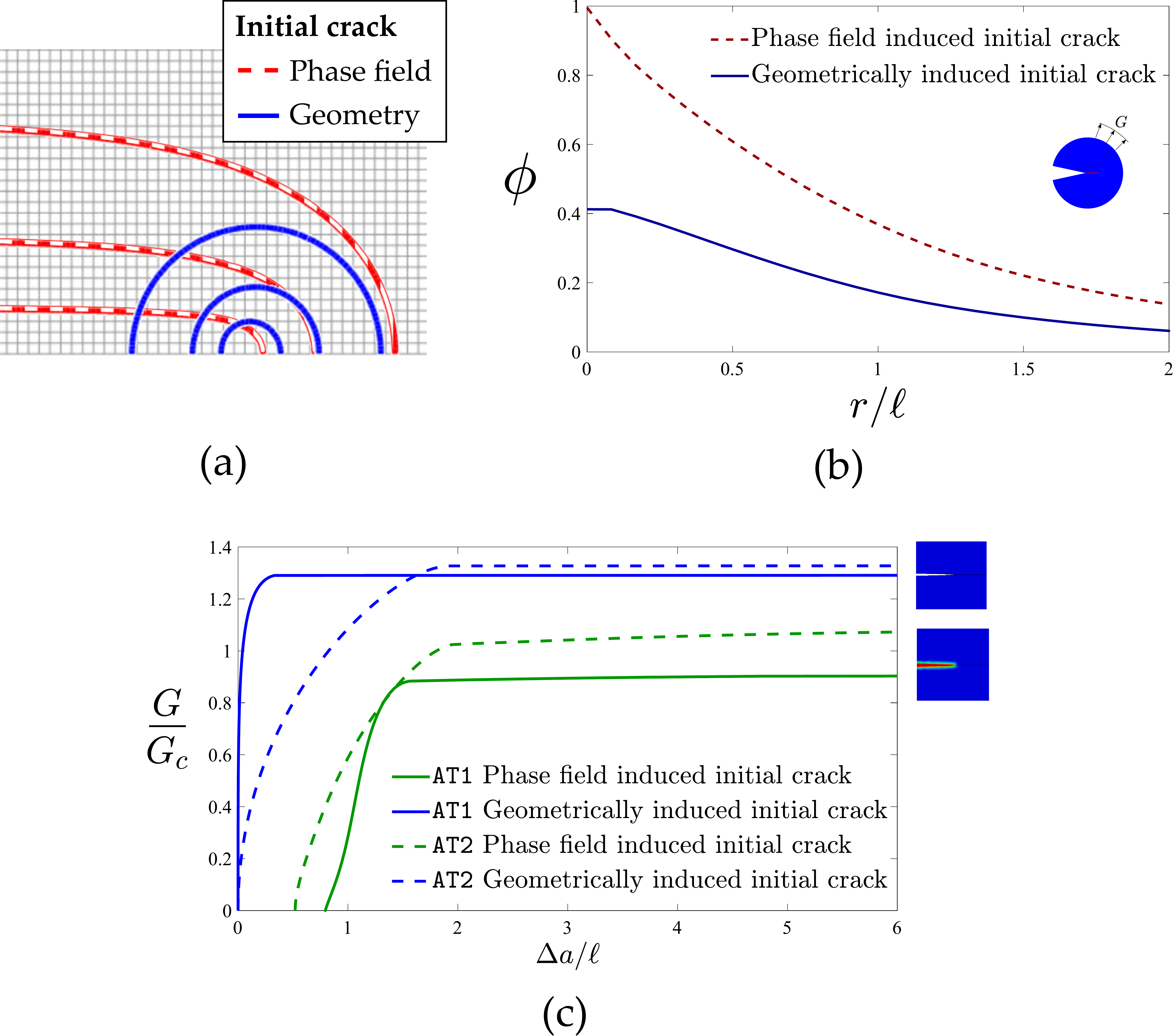}}%
  \cprotect\caption{Boundary layer analysis. (a) Phase field contours for the \verb|AT1| model and both geometrically (red, dashed) and phase field (blue, solid) induced initial cracks; (b) crack tip phase field distribution for the \verb|AT1| model ($r, \theta=0^\circ$), shortly before the onset of crack growth; and (c) crack extension, as computed from the crack surface density function - Eq. (\ref{eq:McMeeking}). The value of the characteristic fracture length scale to the phase field length scale is $L_f/\ell=10$.}
  \label{fig:Contours}
\end{figure}

Secondly, we observe that the predictions of the \verb|AT2| model lead to higher $G$ values than those of the \verb|AT1| crack density function. This could be due to the larger unloading region exhibited in the \verb|AT2| model after the critical stress has been reached, see Fig. \ref{fig:AT1AT2tractionseparation}. Thirdly, and arguably most importantly, while all mesh-converged values of $G$ at crack initiation approach $G_c$, the approximation is notably better when the initial crack has been introduced by prescribing the nodal values of the phase field. Values of $G/G_c$ very close to unity are attained with both \verb|AT1| and \verb|AT2| models when the initial crack has been defined using the phase field, showcasing the agreement between phase field models and classical fracture mechanics theory. However, when the crack is introduced geometrically (e.g., by duplicating the nodes along the crack faces), the magnitude of the energy release rate at the initiation of crack growth is noticeably larger than the fracture energy ($G/G_c \approx 1.3$). We further investigate this by plotting the phase field contours in the vicinity of the crack tip, the distribution of $\phi$ along the extended crack plane ($r, \theta=0^\circ$), and the crack extension as a function of the remote $G$ - see Fig. \ref{fig:Contours}.  

We shall first discuss the phase field contours, Fig. \ref{fig:Contours}a. The qualitative results presented emphasise the additional energy cost associated with prescribing the crack geometrically; the phase field has to increase upon loading in all directions surrounding the crack tip. On the other hand, the phase field is already fully developed at the existing crack faces when the crack is prescribed through the phase field region. This energy barrier can rationalise the differences observed in Fig. \ref{fig:LengthScales}. Another related and relevant effect is the role that the phase field natural boundary condition plays. Upon making use of the constitutive choices, Eq. (\ref{eq:balance_BC}b) can be re-formulated as,
\begin{equation}
    \nabla \phi \cdot \textbf{n} = 0 \, ,
\end{equation}

\noindent implying that the phase field variable $\phi$ must approximate the free surface with a zero slope. This is shown in Fig. \ref{fig:Contours}b, which depicts the distribution of $\phi$ along the extended crack plane ($\theta=0^\circ$), with $r$ being the distance ahead of the crack tip. Results are shown as computed with the \verb|AT1| model and for both the cases of the initial crack being induced by the phase field and geometrically. The distribution of $\phi$ shown corresponds to an instant close to the fracture event; due to the sudden crack extension observed in the case of the geometric initial crack, the phase field variable takes significantly lower values. More importantly, when the crack is induced geometrically a plateau can be observed close to the crack tip, while the result obtained for the phase field induced initial crack (where there is no free crack surface) reveals a monotonically increasing $\phi$ as $r \to 0$. Finally, let us turn our attention to Fig. \ref{fig:Contours}c. The crack extension has been computed using the crack surface expression given in Eq. (\ref{eq:McMeeking}). Thus, the cases where the initial crack has been introduced prescribing the phase field variable $\phi=1$ exhibit a non-zero crack surface even before applying the load, as $\phi$ drops away from the crack tip in a smooth manner, with the smearing of the crack controlled by the magnitude of $\ell$. This additional contribution to the crack surface, not present in the case of the geometrically-induced cracks, is likely to contribute to the different values of $G$ measured at crack initiation. The figure also highlights a key difference between the \verb|AT1| and \verb|AT2| formulations. Due to the lack of a purely elastic phase in \verb|AT2| models, the magnitude of the crack surface formed prior to brittle fracture is larger.\\ 

\begin{figure}[H]
    \centering
    \includegraphics[width=\textwidth]{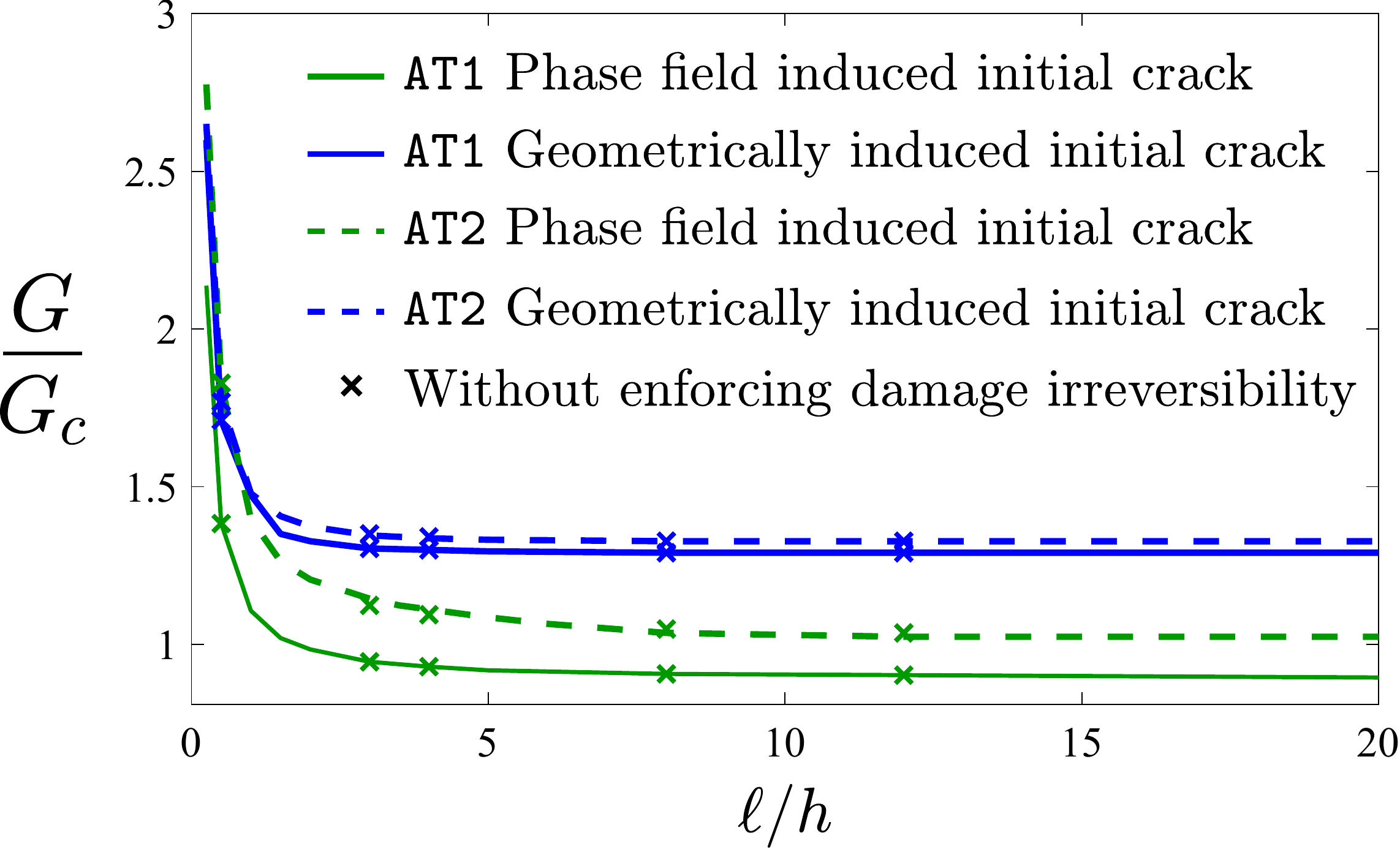}
    \caption{Boundary layer analysis: Influence of enforcing damage irreversibility on the initiation of crack growth. Symbols denote the results without the irreversibility condition, Eq. (\ref{eq:History}). The value of the characteristic fracture length scale to the phase field length scale is $L_f/\ell=10$.}
    \label{fig:MBL_reverse}
\end{figure}

There could be other factors influencing the precision of the phase field fracture model in predicting the initiation of crack growth. One aspect that has been discussed in the literature \cite{Linse2017,Strobl2020} is the influence of the damage irreversibility condition (Section \ref{Sec:DamageIrreversibility}). In particular, it has been argued that the irreversibility condition (\ref{eq:History}) prevents the phase field from attaining its optimal crack profile, providing a source of inaccuracy. Thus, we re-calculate Fig. \ref{fig:LengthScales} enforcing the irreversibility condition (\ref{eq:History}) only when $\phi \ge 0.95$, ensuring damage irreversibility in fully cracked material points but leaving the gradients free to form their optimal profile. As shown in Fig. \ref{fig:MBL_reverse}, no noticeable effect is observed in the present boundary value problem (a long, infinitesimally sharp crack). Differences are in all cases below 0.2\%. 

\begin{figure}[H]
    \centering
    \includegraphics[width=\textwidth]{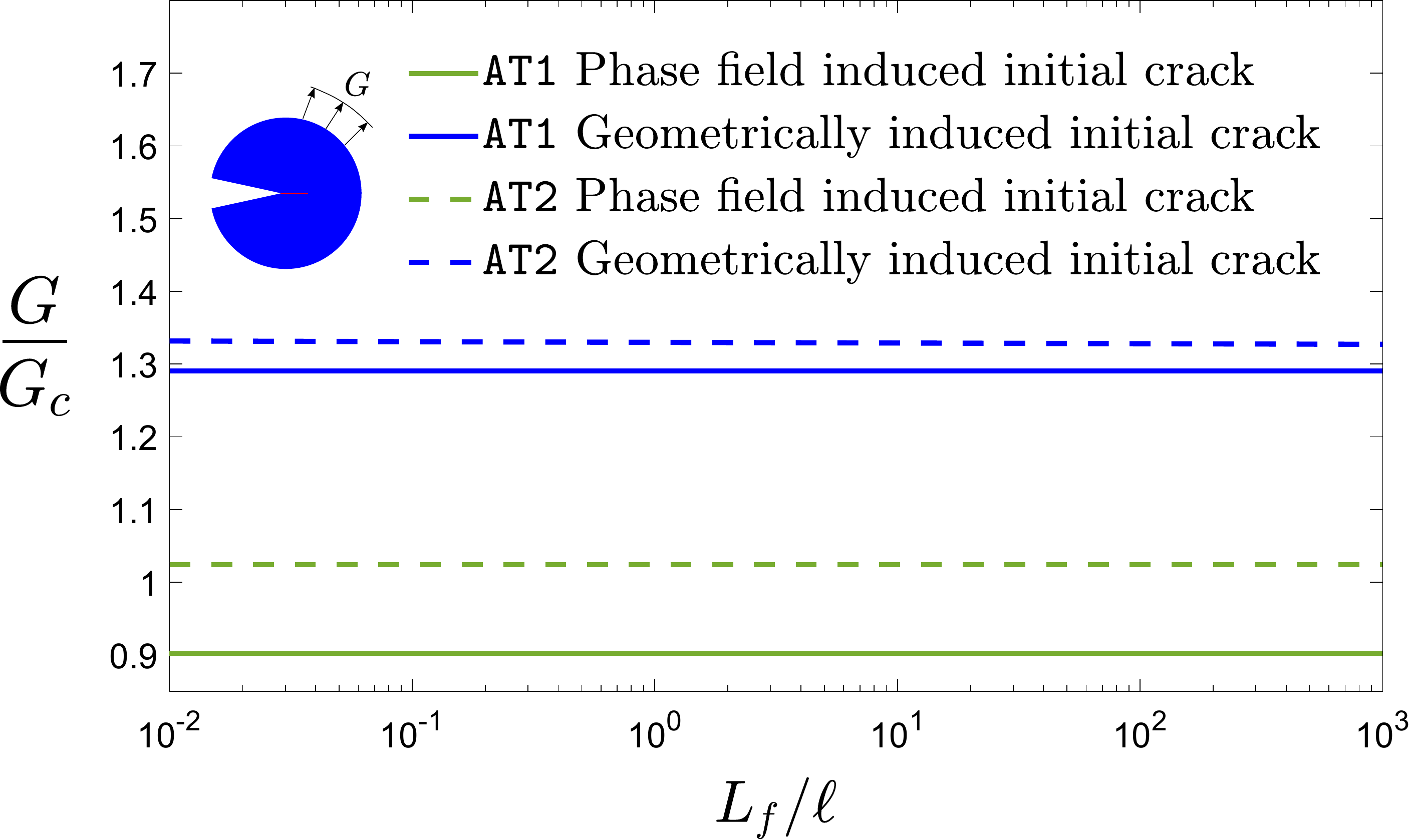}
    \caption{Boundary layer analysis: influence of varying the fracture length scale $L_f=G_c (1-\nu^2)/E$, with $\ell/h=12$.}
    \label{fig:MBL_Lf}
\end{figure}

Finally, we explore the role of the value of the characteristic fracture length scale to the phase field length scale, $L_f/\ell$, as the $\Gamma$-convergence properties of the approximation of the Griffith functional by the phase field functional hold for $\ell \to 0$. The results, shown in Fig. \ref{fig:MBL_Lf}, reveal a negligible influence of the $L_f/\ell$ over a range spanning six orders of magnitude. Thus, the results and conclusions from Fig. \ref{fig:LengthScales} hold; phase field fracture predictions for the initiation of crack growth are close to those of classical fracture mechanics but the approximation improves if the initial crack is defined using the phase field variable.

\subsection{Stable crack growth: double cantilever beam analysis}
\label{Sec:DCB}

We shall now investigate the effectiveness of phase field fracture methods in approximating stable crack growth. For this purpose, we will model crack propagation in a double cantilever beam; a boundary value problem with a known analytical solution, based on beam theory, for relating the energy release rate, the applied displacement and the crack length. To the best of our knowledge, this analysis has not been conducted before.\\

\begin{figure}[H]
    \centering
    \includegraphics[width=\textwidth]{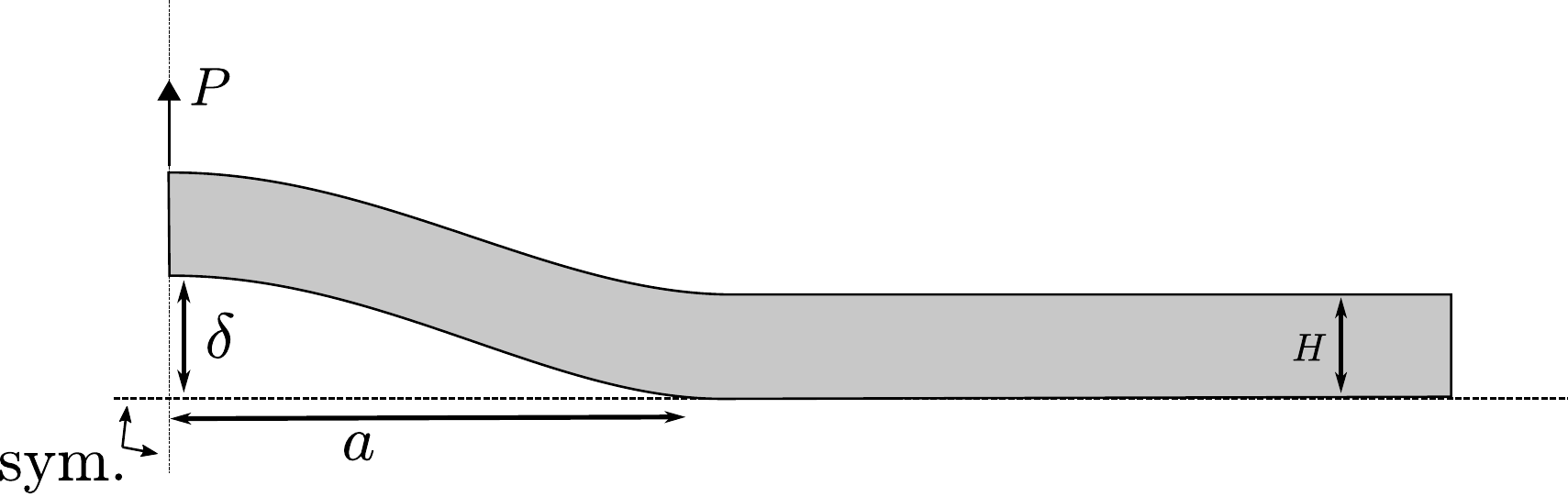}
    \caption{Double cantilever beam analysis: sketch of the boundary value problem. Only one quarter of the problem is modelled taking advantage of symmetry.}
    \label{fig:DCBsketch}
\end{figure}

\begin{figure}[H]
  \makebox[\textwidth][c]{\includegraphics[width=1.2\textwidth]{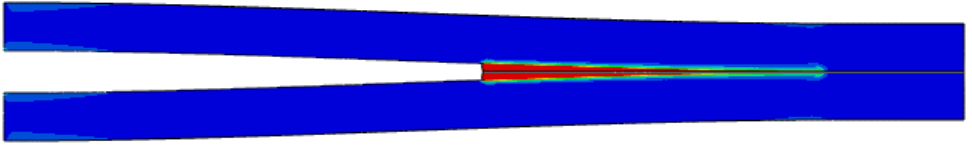}}%
    \cprotect\caption{Double cantilever beam analysis: snapshot of the crack propagation process, with red colour denoting fully cracked material points ($\phi > 0.95$) and blue colour used to denote intact material points ($\phi \approx 0$).}
    \label{fig:DCB_contour}
\end{figure}

The geometry of the model is shown in Fig. \ref{fig:DCBsketch}. Only a quarter of the boundary value problem is modelled, taking advantage of symmetry. Plane strain conditions are assumed, with a thickness of $B=1$ mm. The height is also taken to be equal to $H=0.9$ mm.  The initial crack length is given by $a=a_0=10$ mm and a vertical crack mouth opening displacement $\delta$ is prescribed along the symmetry axis. Here, Poisson's ratio is also taken to be $\nu=0.3$. The model is discretised using a total of 190,140 quadratic quadrilateral elements with reduced integration, with the characteristic length of the element along the crack propagation region being ten times smaller than the phase field length scale (in agreement with the mesh sensitivity analysis conducted above; Fig. \ref{fig:LengthScales}). The phase field length scale is chosen to be $\ell=0.03$ mm and $L_f/\ell\approx 0.003$. A representative result of crack propagation is shown in Fig. \ref{fig:DCB_contour}, where the red colour is employed to denote fully cracked material points ($\phi > 0.95$) while blue colour is used to denote intact material points ($\phi \approx 0$).

An analytical relation between the energy release rate and the applied displacement for a given crack size, $a$, can be readily derived using Timoshenko beam theory. 
The relationship between the transverse force $P$ acting on the quarter model (see Fig. \ref{fig:DCBsketch}) and the displacement $\delta$ is given by
\begin{equation}
    \delta = \dfrac{P a^3 }{\bar EBH^3} + \dfrac{P a }{\kappa\mu BH}\, .
\end{equation}
Here, $\bar E=E/(1-\nu^2)$ is the plane strain Young's modulus, $\mu=E/(2(1+\nu))$ is the shear modulus, and $\kappa\approx 5/6$ is the shear coefficient for the rectangular beam cross section. Exployting symmetry around the horizontal axis, the energy release rate may be calculated from the compliance $C=\delta/P$ by
\begin{equation}
    G= 2\cdot\dfrac{P^2}{2B}\frac{\text{d}C}{\text{d}a}=\dfrac{3P^2a^2}{\bar E B^2H^3}+ \dfrac{P^2 }{\kappa \mu B^2H} \, ,
\end{equation}
Accordingly, the energy release rate can be formulated as a function of displacement, $\delta$, and crack length, $a$, as follows:
\begin{equation}
\label{eq:DCB_analytical}
    G= \dfrac{3\bar EH^3}{a^4}\cdot\dfrac{1+\dfrac{\bar E}{3\kappa\mu}\left(\dfrac{H}{a}\right)^2}{\left(1+\dfrac{\bar E}{\kappa\mu}\left(\dfrac{H}{a}\right)^2\right)^2} \cdot \delta^2\, ,
\end{equation}
and thus, for a given material toughness $G_c$, Eq. (\ref{eq:DCB_analytical}) provides a unique relation between the beam displacement $\delta$ and the crack length $a$.\\

\begin{figure}[H]
    \centering
    \begin{subfigure}[H]{1\textwidth}
        \includegraphics[width=0.9\textwidth]{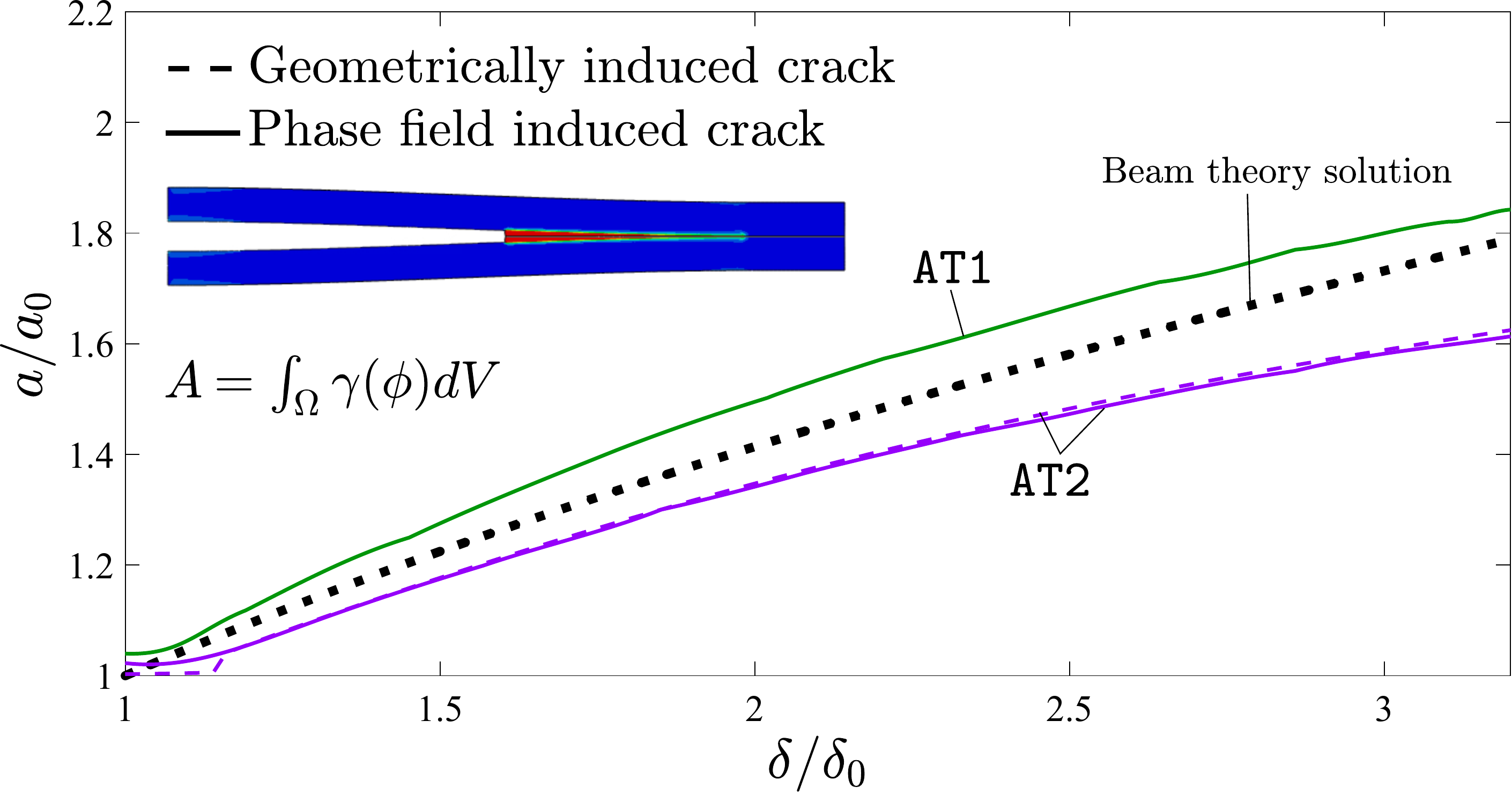}
        \caption{}
        \label{Fig:Sq1}
    \end{subfigure}
    \hspace{1pt}
    \begin{subfigure}[H]{1\textwidth}
        \includegraphics[width=0.9\textwidth]{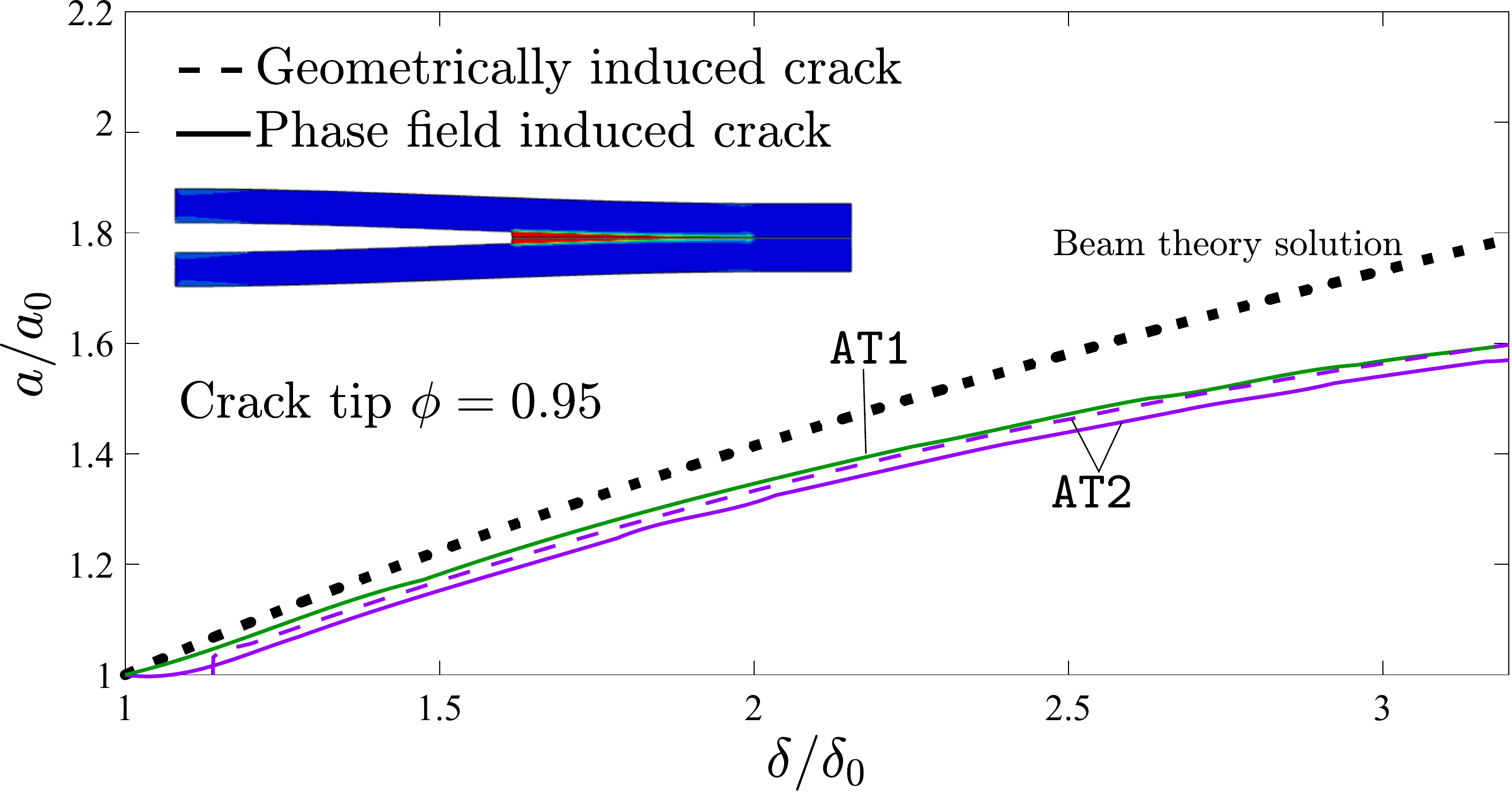}
        \caption{}
        \label{Fig:Sq2}
    \end{subfigure}
    \cprotect\caption{Double cantilever beam analysis. Crack extension as measured by: (a) the crack surface density function, Eq. (\ref{eq:McMeeking}), and (b) the furthest point with $\phi=0.95$.}
    \label{fig:DCB_a}
\end{figure}

As in the previous case study, we employ both the \verb|AT1| and \verb|AT2| models and assess as well the influence of either defining the initial crack geometrically or using the phase field. In addition, unlike the previous analysis, results are now sensitive to the methodology employed to measure crack extension; we choose to compare two options: (1) using the crack surface integral (\ref{eq:McMeeking}), and (2) assuming that the crack front is given by the $\phi=0.95$ contour. The results for each approach are shown in Figs. \ref{fig:DCB_a}a and \ref{fig:DCB_a}b, respectively.\\

A satisfactory agreement is observed. All finite element results provide a stable cracking response that qualitatively mimics that of the analytical solution. However, quantitative differences can be observed, and these are particularly noticeable for specific modelling and constitutive choices. Namely, a better agreement is attained when the crack extension is measured using the crack density function (\ref{eq:McMeeking}), as opposed to assuming the crack front to be the furthest point with $\phi=0.95$. Also, the \verb|AT1| model appears to deliver predictions that are closer to the beam theory solution, relative to the \verb|AT2| model. Nevertheless, all results appear to display a similar shape and the differences are mainly related to the onset of crack initiation and the length of the initial crack. Thus, all phase field results require a larger applied displacement to initiate the fracture process. Also, as in the previous case study, the \verb|AT1| model with a phase field induced crack overpredicts the initial crack length, as there is a contribution from the gradients of $\phi$ to the crack density function. Some general trends to take note of are that the problem exhibits some sensitivity to the size of $\ell/H$ and that all constitutive choices exhibit a slowing of the crack growth relative to the analytical solution. This is seemingly not caused by edge effects, as identical results have been obtained for a beam of length $L=30$ mm. The correspondence between the analytical and predicted curves may be improved by accounting for the slight loss of bending stiffness caused by the degradation from the phase field in the gradient region. Furthermore, the phase field attains non-zero values along the top edge which further reduces the bending stiffness of the beam. The latter may be remedied by introducing a strain split scheme for preventing damage evolution from compression \cite{Amor2009,Miehe2010}.

\subsection{Size effects and the transition flaw size}
\label{Sec:TransitionFlawSize}

So far, we have focused on the original phase field fracture formalism, with the aim of providing a regularisation that accurately approximates Griffith's energy balance in the $\ell \to 0^+$ limit. Thus, the length scale has been considered \emph{exclusively} a regularising parameter. However, there is an increasing interest in investigating the implications of considering a finite phase field length scale and the resulting analogies with gradient damage models \cite{Duda2015,Tanne2018,JMPS2020}. As discussed in Section \ref{sec:Theory}, the consideration of a finite $\ell > 0^+$ introduces a critical stress proportional to $1/\sqrt{\ell}$, which is absent in Griffith's formulation and linear elastic fracture mechanics. Thus, $\ell$ becomes a material property. The motivation for adopting a positive, constant $\ell$ stems from the fact that Griffith's theory is unable to capture some well-characterised size effects. One of these important size effects is the transition flaw size concept, which is the cornerstone of many engineering standards and fracture mechanics-based engineering design; if a crack is smaller than the transition flaw size, then the crack will not grow and the specimen will fail at the material strength (or at the yield stress $\sigma_y$, if plastic design is considered). We shall show here that the transition flaw size paradigm is a natural byproduct of variational phase field fracture models that consider $\ell$ to be an internal material length.\\

\begin{figure}[H]
    \centering
    \includegraphics[width=0.6\textwidth]{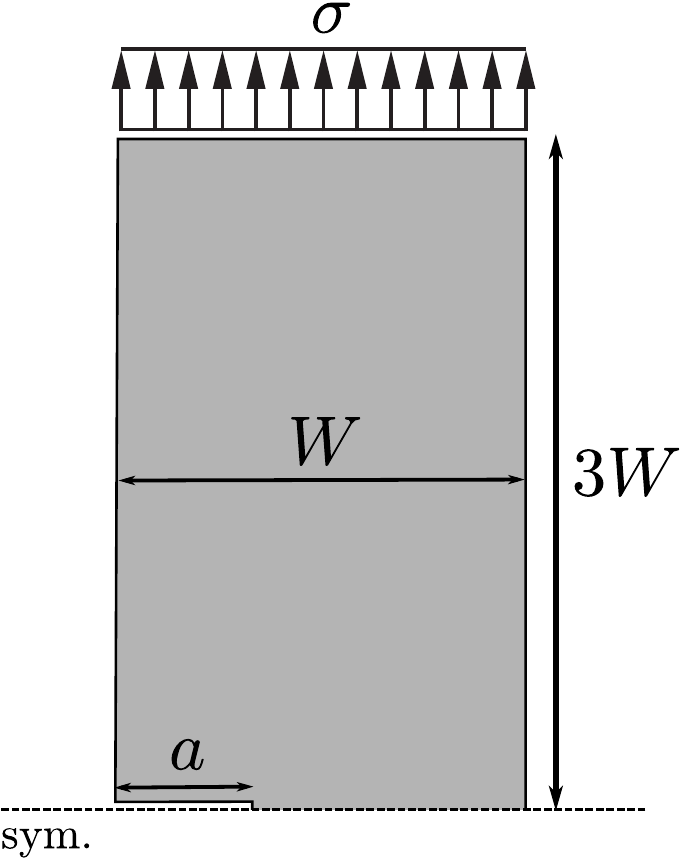}
    \caption{Transition flaw size analysis: geometry and loading configuration of the numerical model.}
    \label{fig:CrackLengthSketch}
\end{figure}

We model fracture in a single-edge notched specimen of width $W$ and height $6W$. The plate is subjected to a remote tensile stress $\sigma$. As shown in Fig. \ref{fig:CrackLengthSketch}, only the upper half of the sample is considered for the finite element analysis, taking advantage of symmetry. The specimen contains a crack of length $a$, which will be varied throughout the analysis. In all cases, the crack is introduced into the model by defining the initial condition $\phi=1$ on the phase field, in agreement with our findings above for best practice. Both the \verb|AT1| and \verb|AT2| models are considered, to assess the implications of different constitutive choices for the crack density function. The phase field length scale is chosen to be small relative to the sample dimensions, $\ell/W=0.03$, and the mesh is refined along the crack ligament, where the characteristic element length equals $h / \ell=0.1$. The model is discretised using a total of 11,251 quadratic quadrilateral elements with reduced integration. 

\begin{figure}[H]
    \centering
    \begin{subfigure}[H]{1\textwidth}
        \includegraphics[width=0.9\textwidth]{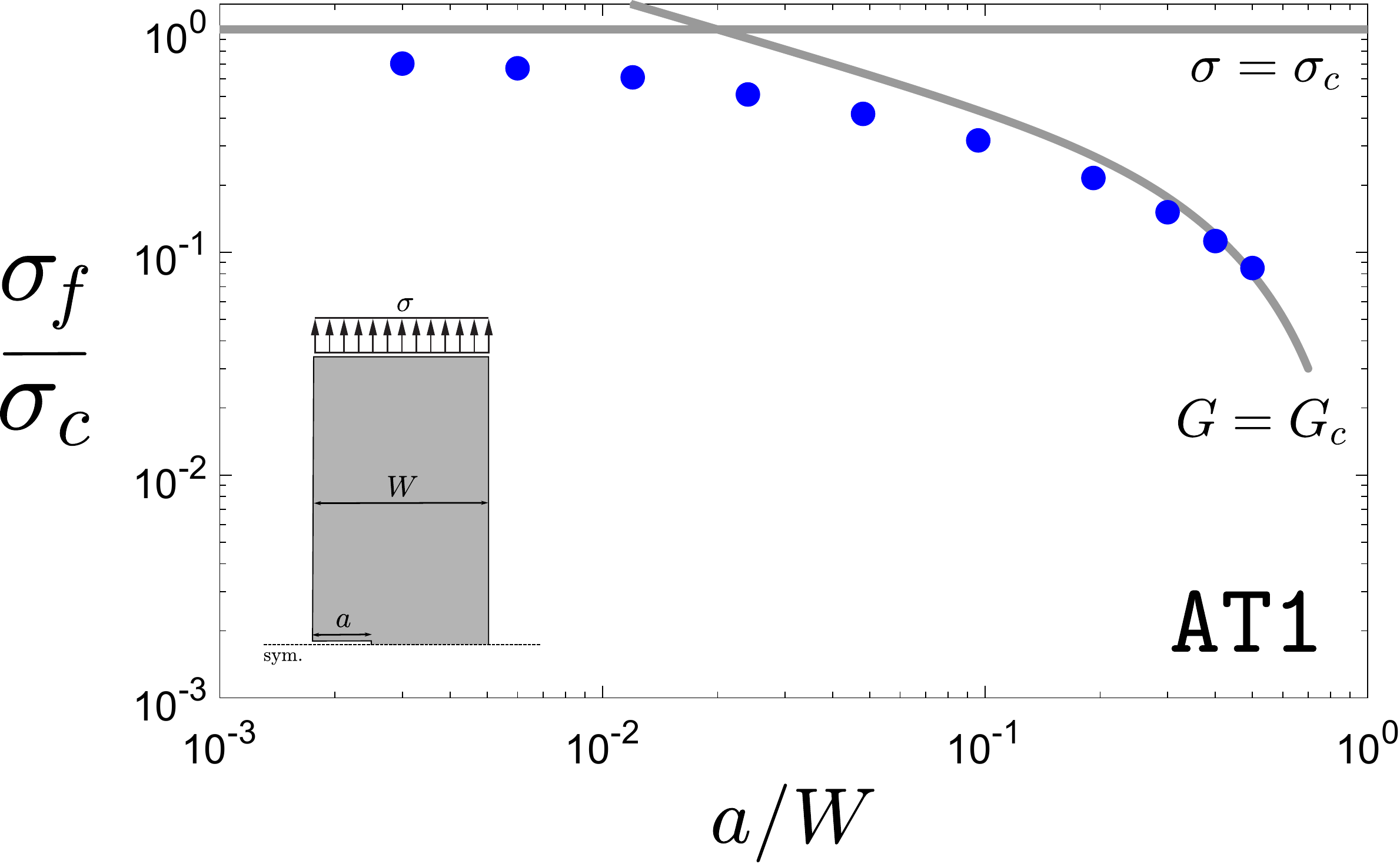}
        \caption{}
        \label{Fig:Sq-geo}
    \end{subfigure}
    \hspace{1pt}
    \begin{subfigure}[H]{1\textwidth}
        \includegraphics[width=0.9\textwidth]{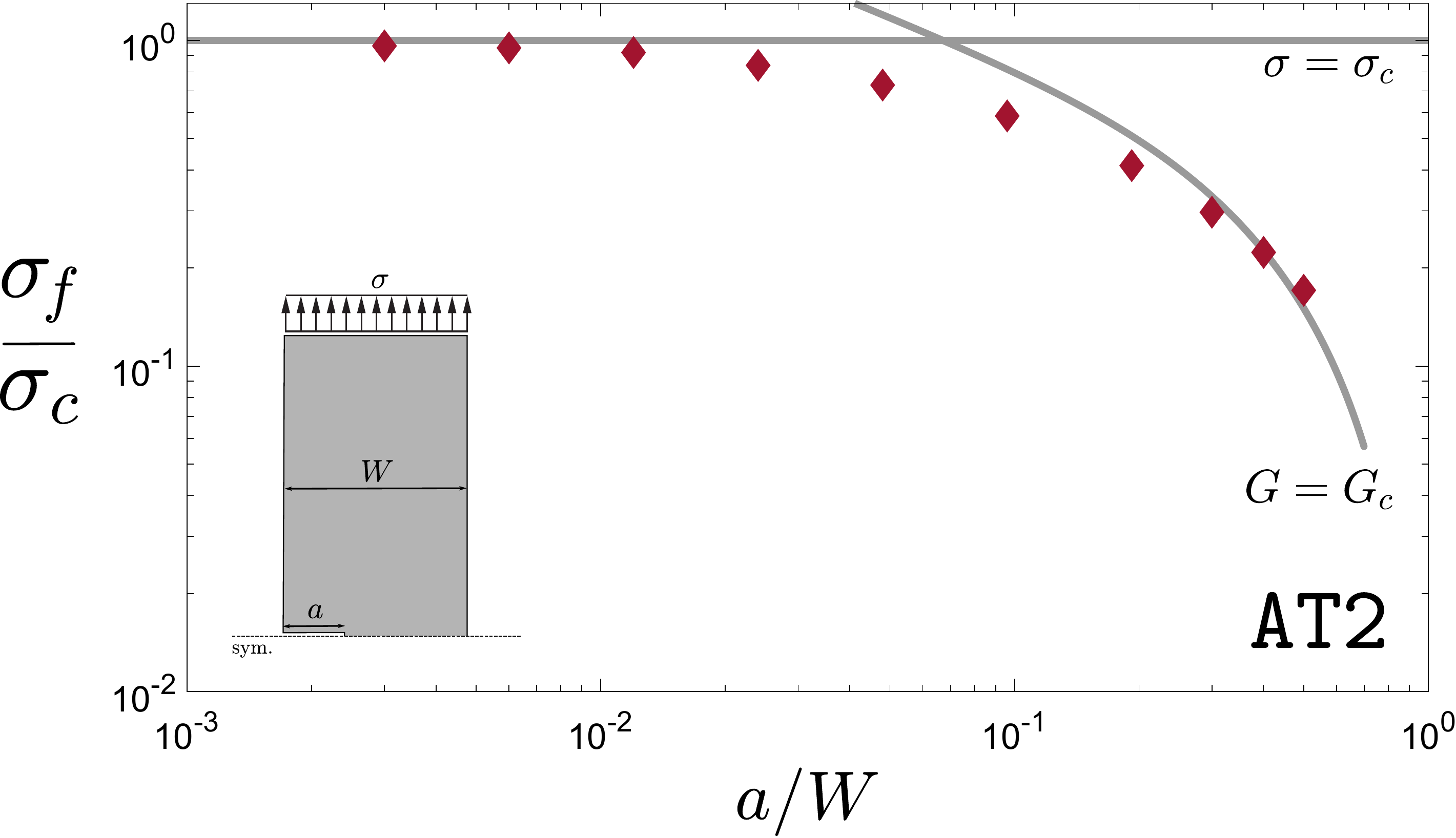}
        \caption{}
        \label{Fig:Sq-phi}
    \end{subfigure}
    \cprotect\caption{Transition flaw size analysis: Failure strength as a function of the crack size for (a) \verb|AT1|, and (b) \verb|AT2| phase field models. The solid grey lines denote the fracture predictions according to the material strength and to Griffith's criterion (\ref{eq:GriffithPlate}).}
    \label{fig:CrackLength}
\end{figure}

The results obtained are shown in Fig. \ref{fig:CrackLength}, where we have superimposed the strength failure criterion (also referred to as \emph{plastic collapse} if $\sigma_c=\sigma_y$) and the Griffith (linear elastic fracture mechanics) prediction:
\begin{equation}\label{eq:GriffithPlate}
    \sigma = \sqrt{\frac{E G_c}{\pi a \left( 1 - \nu^2 \right) }} \frac{1}{f \left( a / W \right)} \, ,
\end{equation}

\noindent with the following geometry factor $f \left( a / W \right)$ for a plate of finite size with an edge crack:
\begin{align}
    f\left(\frac{a}{W}\right) = & \left(\dfrac{2W}{\pi a}\tan\dfrac{\pi a}{2W}\right)^{1/2}\left(\cos\dfrac{\pi a}{2 W}\right)^{-1} \\ \nonumber 
    & \left[ 0.752+2.02\dfrac{a}{W}+0.37\left(1-\sin\dfrac{\pi a}{2 W}\right)^3\right] \, .
\end{align}

The results shown in Fig. \ref{fig:CrackLength} are given in terms of the failure stress $\sigma_f$ as a function of the crack size $a$, for both \verb|AT1| and \verb|AT2| models. Note that the material strength takes different values for each of these constitutive choices - see (\ref{eq:sigmacAT1AT2}). It can be observed that phase field fracture models are capable of reconciliating stress and toughness criteria for fracture; a good agreement with the Griffith criterion is observed for large cracks and predictions transition smoothly to a strength-driven failure as the crack size decreases below the transition flaw size.

In terms of constitutive choices for the crack density function, both \verb|AT1| and \verb|AT2| models appear to provide a good agreement with the limiting cases of $\sigma_f=\sigma_c$ and $G=G_c$. The agreement appears to be slightly better for the \verb|AT2| case, in that the material strength is only attained when using the \verb|AT1| model for very short cracks (much smaller than the transition flaw size). Or, in other words, the transition between pure strength and pure toughness-driven criteria appears to span a wider range of crack sizes in the case of the \verb|AT1| formulation.

\section{Discussion}
\label{Sec:Discussion}

Our results show that phase field fracture models provide a good approximation to classical fracture mechanics predictions. Three research questions have been answered, the first one being: can phase field fracture methods capture crack growth initiation at the appropriate energy release rate? By modelling crack growth from an existing (long) crack upon the application of a remote $G$ (or $K$), we have seen that cracking takes place at $G \approx G_c$ but only for certain modelling choices. Specifically, using the phase field variable to induce the initial crack provides a result closer to Griffith's criterion, while introducing the crack geometrically leads to crack initiation values of $G$ that are slightly larger than $G_c$. It is important to emphasise that this finding relates to a long, infinitesimally-sharp crack. Similar conclusions were attained by Klinsmann \textit{et al.} \cite{Klinsmann2015} for a finite crack using the \verb|AT2| model and a pure bending boundary value problem. We also find that the specific constitutive choice for the crack density function (\verb|AT1| vs \verb|AT2|) does not play a significant role, with the \verb|AT2| model predicting the initiation of crack growth at a slightly larger $G$. We have also shown that these conclusions hold for $\ell \to 0$, independently of whether or not the irreversibility condition is employed. In fact, enforcing damage irreversibility appears to have a negligible effect on the efficacy of phase field models for accurately predicting crack initiation, under the conditions considered here (a long sharp crack). Our analysis suggests that the delay in initiating fracture when the initial crack is geometrically prescribed is related to the natural boundary condition (\ref{eq:balance_BC})b, constraining $\phi$ to be constant near the crack surface, and the additional energy expenditure required to build-up a highly constrained phase field region around the crack tip (even behind the crack). There are other sources that can potentially contribute to discrepancies in crack initiation predictions, which have not been quantified as we have judged them \textit{a priori} to be of secondary importance. For example, in elastic-plastic solids, plasticity introduces non-proportional straining but, while this effect can be significant during continued crack growth \cite{JAM2018,EFM2019}, a very minor influence is expected for crack initiation under small scale yielding conditions. Also, the extent to which a Griffith-like energy balance can be used for ductile solids is questionable \cite{Orowan1948,Gurtin1979,Hutchinson1983,Stevens1991,Duda2015}; the thermodynamics picture will change, as local plastic flow provides a localised source of heat. Along the same lines, different constitutive choices for the crack density function can also provide different degrees of approximation, in the same way that this is observed with different traction-separation laws in cohesive zone models, where larger unloading regimes in the traction-separation law lead to larger differences compared to a proportional loading scenario. Hence, for cohesive zone models, a trapezoidal law with a smaller unloading region, like the one by Tvergaard and Hutchinson \cite{Tvergaard1992}, can provide a better approximation of crack initiation, relative to an exponential law, such as that by Xu and Needleman \cite{Xu1994}, with a large unloading regime. In any case, differences are expected to be small also for phase field models; our results show that both \verb|AT1| and \verb|AT2| models predict the initiation of crack growth at $G \approx G_c$ (for a phase field induced initial crack) despite their different unloading regimes - see Fig. \ref{fig:AT1AT2tractionseparation}. It must be emphasised that our findings are related to solids containing cracks; phase field fracture models can also predict the nucleation of cracks from pristine samples and non-sharp defects such as notches, where the conclusions reported here might not apply. In particular, the conclusions drawn in regard to the irreversibility condition might change \cite{Linse2017,Strobl2020} and it has been reported that prescribing $\phi=1$ at the defect surface is not the most accurate way of capturing crack nucleation from blunted notches \cite{Tanne2018}.\\

The second research question deals with the capabilities of phase field fracture models in predicting stable crack growth, in agreement with beam theory and the fracture energy balance. We gained new insight by modelling the progressive failure of a double cantilever beam with a known analytical solution, based on Euler-Bernoulli beam theory. The results revealed a satisfactory agreement but also noticeable quantitative differences depending on the approach employed to measure the crack extension and the constitutive model. The best result was attained by employing the \verb|AT1| model and, more importantly, measuring the degree of crack extension through the crack density function - Eq. (\ref{eq:McMeeking}). The lack of a similar study in the literature, to the best of our knowledge, hinders gaining further insight by comparing to previous studies.\\

Finally, we aimed at shedding light on the capabilities of phase field fracture models to capture, by attributing $\ell$ a physical meaning, well-known size effects that cannot be predicted with Griffith theory. Griffith's framework and linear elastic fracture mechanics can capture how the critical load for fracture scales as $1/\sqrt{L}$, where $L$ is the reference size of the specimen, and how the strength of the specimen decreases with increasing crack size. However, this scaling size effect breaks down as the load required to fracture small samples ($L \to 0$) does not go to infinity -  cracks do not propagate if they are smaller than a reference length (the transition flaw size), and failure by other mechanisms sets in. These inconsistencies can be addressed by incorporating a length scale or a critical strength. In the context of phase field models, the consideration of $\ell$ as a material constant naturally introduces a critical stress - see (\ref{eq:CriticalStress}). We have shown that this approach can readily capture the transition flaw size concept, gradually changing from toughness-driven to strength-driven failures. This is observed with the initial crack prescribed using the phase field and for both \verb|AT1| and \verb|AT2| models (with a slightly better performance using the latter). Similar conclusions were drawn by Tann\'{e} \textit{et al.} \cite{Tanne2018} using a different boundary value problem (a plate with a central crack), the \verb|AT1| model and a geometrically-induced initial crack. Thus, our findings demonstrate that phase field models without an elastic phase can also reconcile toughness and strength. The capabilities of variational phase field models in incorporating the concepts of material strength and toughness bring them in agreement with the coupled criterion of finite fracture mechanics \cite{Leguillon2002,Molnar2020a}, but with the additional modelling capabilities intrinsic to phase field models. Along these lines, several modelling strategies and constitutive prescriptions have been presented to enhance the crack nucleation capabilities of phase field models and decouple the strength and the phase field length scale \cite{Wu2017,Sargado2018,Kumar2020a}.

\section{Conclusions}
\label{Sec:Conclusions}

We have reviewed the most widely used phase field fracture models and revisited their ability to deliver predictions in agreement with classical fracture mechanics theory. The energy balance of Griffith theory was cast in a variational form and approximated using a regularised phase field functional. Then, the nucleation and growth of cracks were predicted based on this global energy minimisation problem. We focused our efforts on three boundary value problems of particular relevance, all of which involve solids containing sharp cracks.\\

First, we used a boundary layer model to impose an increasing $G$ and assess whether phase field fracture can predict the initiation of growth at $G=G_c$. We found that this result is only attained with accuracy if the initial crack is introduced by prescribing the initial value of the phase field variable $\phi$, while the models containing a geometrically-induced crack overestimate the critical value of $G$. From our predictions of phase field distribution and crack surface evolution, we conclude that this is due to the natural boundary condition for the phase field $\nabla \phi \cdot \textbf{n} = 0$ and the energy barrier associated with the build-up of a highly constrained phase field region around the crack tip. In addition, we have tested and discussed other hypotheses that can potentially rationalise the mismatch with Griffith's criterion; we conclude that, for the conditions considered here, the irreversibility condition and non-proportional straining play a secondary role. Secondly, we assessed for the first time the capabilities of phase field fracture in predicting sustained, stable crack growth in agreement with beam theory and Griffith's energy balance. While all predictions were deemed satisfactory, the degree of agreement improved notably if the crack extension was measured using the crack density functional and, for the \verb|AT1| model, if the crack was introduced using the phase field. Finally, we treated the phase field length scale as a material property and modelled the failure of a plate with different crack sizes, showing that the consideration of a constant $\ell>0^+$ enables capturing the vanishing effect of small flaws on the fracture strength and reconciles toughness and strength failure criteria. This size effect, which cannot be captured by Griffith's theory, was appropriately predicted with both \verb|AT1| and \verb|AT2| models.\\

It is therefore concluded that phase field models can deliver accurate fracture predictions if suitable modelling choices are made. Specifically, we note that (for the conditions examined here) the constitutive choices for the crack density function (\verb|AT1| vs \verb|AT2|) play a secondary role but accuracy can be improved noticeably if the initial crack is defined using the phase field and the crack extension is measured using the crack density function (\ref{eq:McMeeking}). These findings have been discussed in the context of the literature, emphasising the new and complementary insight provided, which is hoped to be valuable in assessing the capabilities of phase field fracture models in delivering predictions in agreement with the energy balance that gave birth to fracture mechanics. 

\section{Acknowledgments}
\label{Sec:Acknowledge of funding}

The authors gratefully acknowledge financial support from the Danish Hydrocarbon Research and Technology Centre (DHRTC). E. Mart\'{\i}nez-Pa\~neda additionally acknowledges financial support from the EPSRC (grants EP/R010161/1 and EP/R017727/1) and from the Royal Commission for the 1851 Exhibition (RF496/2018).

\bibliographystyle{elsarticle-num} 
\bibliography{library}

\end{document}